\documentclass[12pt,notitlepage]{article}
\usepackage{amsmath,amssymb,amsthm}
\usepackage{cite}
\usepackage{graphics}
\usepackage{latexsym}
\usepackage{amsmath}
\usepackage{amscd}
\usepackage{graphicx}
\begin{document}
\begin{center}{\Large \bf The bi-Hamiltonian structure and new solutions of KdV6 equation}
\end{center}
\begin{center}
{ Yuqin Yao\footnote{yqyao@math.tsinghua.edu.cn} and Yunbo
Zeng\footnote{yzeng@math.tsinghua.edu.cn}  }
\end{center}

\begin{center}{{\small \it Department of Mathematics,
Tsinghua University, Beijing 100084 , PR China}}
\end{center}
\vskip 12pt { \small\noindent\bf Abstract.}
 {We show that the KdV6 equation recently studied in \cite{kdv6,k}
 is equivalent to the Rosochatius deformation of KdV equation with
 self-consistent sources (RD-KdVESCS) recently presented in
 \cite{y}. The $t$-type bi-Hamiltonian formalism of KdV6 equation
 (RD-KdVESCS) is constructed by taking $x$ as evolution parameter.
 Some new solutions of KdV6 equation, such as soliton, positon and negaton solution, are presented.
  }

\section{Introduction}
Recently, the 5 authors of \cite{kdv6} applied the
Painlev$\acute{e}$ analysis to the class of sixth-order nonlinear
wave equations, and found 4 cases that pass the Painlev$\acute{e}$
test. Three of those cases correspond to previously known integrable
equations, whereas the fourth one turns out to be new:
\begin{equation}
\label{eqns:1}
(\partial_{x}^{3}+8u_{x}\partial_{x}+4u_{xx})(u_{t}+u_{xxx}+6u_{x}^{2})=0.
\end{equation}
This equation, as it stands, does not belong to any recognizable
theory. In the variables $v=u_{x},~w=u_{t}+u_{xxx}+6u_{x}^{2}$,
(\ref{eqns:1}) is converted to
\begin{subequations}
\label{eqns:2}
 \begin{align}
&v_{t}+v_{xxx}+12vv_{x}-w_{x}=0,\\
& w_{xxx}+8vw_{x}+4w v_{x}=0,
 \end{align}
\end{subequations}
which is referred as KdV6 equation in \cite{kdv6} and regarded as a
nonholonomic deformation of the KdV equation. The authors of
\cite{kdv6} found Lax pair and an auto-B$\ddot{a}$cklund
transformation for (\ref{eqns:2}). They claimed that  (\ref{eqns:2})
is different from the KdV equation with self-consistent sources
(KdVESCS) and reported that they were unable to find higher
symmetries and asked if higher conserved densities and a Hamiltonian
formalism exist for (\ref{eqns:2}).

In \cite{k}, Kupershmidt described (\ref{eqns:2}) as a nonholonomic
perturbations of bi-Hamiltonian systems. By rescaling $v$ and $t$ in
(\ref{eqns:2}), one gets
\begin{subequations}
\label{eqns:3}
 \begin{align}
&u_{t}=6uu_{x}+u_{xxx}-w_{x},\\
& w_{xxx}+4uw_{x}+2w u_{x}=0,
 \end{align}
\end{subequations}
which can be converted into a nonholonomic perturbations of
bi-Hamiltonian systems \cite{k}
\begin{subequations}
\label{eqns:hh}
 \begin{align}
 &u_{t}=B^{1}(\frac{\delta H_{n+1}}{\delta u})-B^{1}(w)=B^{2}(\frac{\delta H_{n}}{\delta
 u})-B^{1}(w),\\
& B^{2}(w)=0,
 \end{align}
\end{subequations}
where
\begin{equation}
\label{eqns:ho}
B^{1}=\partial=\partial_{x},~B^{2}=\partial^{3}+2(u\partial+\partial
u)
\end{equation}
are the two standard Hamiltonian operators of the KdV hierarchy,
$n=2$, and
$$H_{1}=u,~H_{2}=\frac{u^{2}}{2},~H_{3}=\frac{u^{3}}{3}-\frac{u_{x}^{2}}{2},\cdots$$
Then the author in \cite{k} believed that he could prove the
integrability of KdV6 equation by constructing the infinite
commuting hierarchy KdV$_{n}$6 (\ref{eqns:hh}) with a common
infinite set of conserved densities. Some solutions for
(\ref{eqns:2}) were obtained in \cite{kdv6,3}.

The soliton equations with self-consistent sources (SESCS) have
attracted much attention (see \cite{scs}-\cite{12}) and have
important physical applications, for example, the KdV equation with
self-consistent sources (KdVESCS) describes the interaction of long
and short capillary-gravity waves\cite{scs}. The Rosochatius
deformation of finite-dimensional integrable Hamiltonian system
(FDIHS) also has important physical application, for example, the
Garner-Rosochatius system can be used to solve the multicomponent
coupled nonlinear Schr$\ddot{o}$dinger equation\cite{pl}. We
generalized the Rosochatius deformation from FDIHS to SESCS and
presented many Rosochatius deformations of SESCS (RD-SESCS) in
\cite{y}, such as RD-KdVESCS which stationary reduction gives rise
to the well-known generalized Henon-Heiles system\cite{m}.

In this paper, we would like to answer the questions mentioned in
\cite{kdv6}. We will first show that (\ref{eqns:3}) is equivalent to
the Rosochatius deformation of KdVESCS (RD-KdVESCS) presented in
\cite{y}. It is known \cite{6,7} that some soliton equations have
both $x-$ and $t-$ type Hamiltonian formulation. However the
Hamiltonian formulation for KdV6 equation (RD-KdVESCS) can not be
written in usual way. We will  formulate it as an
infinite-dimensional integrable bi-Hamiltonian system with a $t-$
type Hamiltonian operator by taking $t$ as the 'spatial' variable
and $x$ as the evolution parameter as in the case of
KdVESCS\cite{8,9}. Since the KdV6 equation can be regarded as the
KdV equation with no-homogeneous term and $w$ is related to the
square of  eigenfunction, we may apply the method of variant of
constant to find some new solutions of KdV6 equation starting from
the known solutions of KdV equation.

The  present paper is organized as follows. We will first convert
the KdV6 equation into RD-KdVESCS, and present extension of KdV6
equation in section 2. In section 3, we will describe RD-KdVESCS
(KdV6 equation) and RD-mKdVESCS as a $t-$ type Hamiltonian system by
taking $x$ as the evolution parameter, respectively. Then following
the procedure given in \cite{6}-\cite{9} by means of the $t-$ type
Miura transformation relating these two Hamiltonian systems, we will
construct the second $t-$ type Hamiltonian structure for KdV6
equation (RD-KdVESCS) from the first Hamiltonian structure of
RD-mKdVESCS, and present infinite chain of local commuting vector
fields for KdV6 equation. Finally in section 4, starting from the
solutions of KdV, we  obtain many new solutions of KdV6 equation,
such as soliton,  positon and  negaton solution.

\section{\bf KdV6 equation is equivalent to  RD-KdVESCS}
By rescaling $u$ and $t$ and using the Galilean invariance of KdV
equation, KdV6 equation (\ref{eqns:3}) can be rewritten as
\begin{subequations}
\label{eqns:nkdv6}
 \begin{align}
&u_{t}=\frac{1}{4}(u_{xxx}+6uu_{x})-w_{x},\\
& w_{xxx}+4(u-\lambda_{1})w_{x}+2w u_{x}=0
 \end{align}
\end{subequations}
where $\lambda_{1}$ is a parameter.

Set \begin{equation} \label{eqns:wevarphi}
 w=\varphi^{2},
\end{equation}
then (6b) yields
$$w_{xxx}+4(u-\lambda_{1})w_{x}+2wu_{x}=2\varphi[\varphi_{xx}+(u-\lambda_{1})\varphi]_{x}+6\varphi_{x}[\varphi_{xx}+(u-\lambda_{1})\varphi]=0,$$
which immediately gives rise to
$$\varphi_{xx}+(u-\lambda_{1})\varphi=\frac{\mu}{\varphi^{3}},$$
where $\mu$ is an integrable constant. So KdV6 equation
(\ref{eqns:nkdv6}) is equivalent to
\begin{subequations}
\label{eqns:rdkdv6}
 \begin{align}
&u_{t}=\frac{1}{4}(u_{xxx}+6uu_{x})-(\varphi^{2})_{x},\\
& \varphi_{xx}+(u-\lambda_{1})\varphi=\frac{\mu}{\varphi^{3}},
 \end{align}
\end{subequations}
which is just the RD-KdVESCS presented in \cite{y}. The Lax pair for
(\ref{eqns:rdkdv6}) reads\cite{y}
\begin{subequations}
\label{eqns:lax}
 \begin{align}
& \left(\begin{array}{c}
\psi_{1}\\
\psi_{2}\\
 \end{array}\right)_{x}=U\left(\begin{array}{c}
\psi_{1}\\
\psi_{2}\\
 \end{array}\right),~~U=\left(\begin{array}{cc}
0 & 1\\
\lambda-u & 0\\
 \end{array}\right)\\ \nonumber
& \left(\begin{array}{c}
\psi_{1}\\
\psi_{2}\\
 \end{array}\right)_{t}=N\left(\begin{array}{c}
\psi_{1}\\
\psi_{2}\\
 \end{array}\right),\\
 & N=\left(\begin{array}{cc}
-\frac{u_{x}}{4}& \lambda+\frac{u}{2}\\
\lambda^{2}-\frac{u}{2}\lambda-\frac{u_{xx}}{4}-\frac{u^{2}}{2}+\frac{1}{2}\varphi^{2} & \frac{u_{x}}{4}\\
 \end{array}\right)-\frac{1}{2}\frac{1}{\lambda-\lambda_{1}}\left(\begin{array}{cc}
\varphi\varphi_{x}& -\varphi^{2}\\
\varphi_{x}^{2}+\frac{\mu}{\varphi^{2}}& -\varphi\varphi_{x}\\
 \end{array}\right).
 \end{align}
\end{subequations}
More generally, the multi-component extension of KdV6 equation is
given by
\begin{subequations}
\label{eqns:exkdv6}
 \begin{align}
&u_{t}=\frac{1}{4}(u_{xxx}+6uu_{x})-\sum\limits_{j=1}^{N}w_{jx},\\
& w_{jxx}+4(u-\lambda_{j})w_{jx}+2u_{x}w_{j}=0, ~j=1,2,\cdots,\cdots
N.
 \end{align}
\end{subequations}
Under the transformation $w_{j}=\varphi_{j}^{2}$,
(\ref{eqns:exkdv6}) can be converted into the following RD-KdVESCS
\begin{subequations}
\label{eqns:rdexkdv6}
 \begin{align}
&u_{t}=\frac{1}{4}(u_{xxx}+6uu_{x})-\sum\limits_{j=1}^{N}(\varphi_{j}^{2})_{x},\\
&
\varphi_{jxx}+(u-\lambda_{j})\varphi_{j}=\frac{\mu_{j}}{\varphi_{j}^{3}},
~j=1,2,\cdots,N.
 \end{align}
\end{subequations}
The Lax pair for (\ref{eqns:rdexkdv6}) is given by (9a) with
\begin{equation} \label{eqns:n}
 N=\left(\begin{array}{cc}
-\frac{u_{x}}{4}& -\lambda+\frac{u}{2}\\
-\lambda^{2}-\frac{u}{2}\lambda-\frac{u_{xx}}{4}-\frac{u^{2}}{2}+\frac{1}{2}\sum\limits_{j=1}^{N}
\varphi_{j}^{2} & \frac{u_{x}}{4}\\
 \end{array}\right)-\frac{1}{2}\sum\limits_{j=1}^{N}\frac{1}{\lambda-\lambda_{j}}\left(\begin{array}{cc}
\varphi_{j}\varphi_{jx}& -\varphi_{j}^{2}\\
\varphi_{jx}^{2}+\frac{\mu}{\varphi_{j}^{2}}& -\varphi_{j}\varphi_{jx}\\
 \end{array}\right).
\end{equation}

\section{\bf Bi-Hamiltonian structure of KdV6 equation}
In this section, we will follow the method in \cite{6}-\cite{9} to
construct the bi-Hamiltonian formalism for KdV6 equation
(RD-KdVESCS). First we will present the $t-$ type Hamiltonian
formalism for RD-KdVESCS and RD-mKdVESCS. For the RD-KdVESCS
(\ref{eqns:rdkdv6}), set
$$\frac{1}{4}u_{xx}+\frac{3}{4}u^{2}-\varphi^{2}=c,~q_{t}=c_{x}, $$
\begin{equation} \label{eqns:set}
q=u,~p=-\frac{1}{8}u_{x},~Q=\varphi,~P=\varphi_{x},~R=(Q,q,P,p,c)^{T},
\end{equation}
then (\ref{eqns:rdkdv6}) becomes $x-$ evolution equations and can be
written as a $t-$ type Hamiltonian system
\begin{subequations}
\label{eqns:kdvh}
 \begin{align}
&R_{x}=\left(\begin{array}{c}
P\\
-8p\\
(\lambda_{1}-q)Q+\frac{\mu}{Q^{3}}\\
\frac{3}{8}q^{2}-\frac{1}{2}Q^{2}-\frac{1}{2}c\\q_{t}
 \end{array}\right)=K_{1}=\Pi_{0}\nabla H_{1},\\ \nonumber
& where~ \nabla~means~variational~derivative,~\nabla H=(\frac{\delta
H}{\delta Q},\frac{\delta H}{\delta q},\frac{\delta H}{\delta
P},\frac{\delta H}{\delta p},\frac{\delta H}{\delta c})^{T}, ~and
~the~\\ \nonumber &  t-type~ Poisson ~operator ~\Pi_{0}~and
~conserved~ density~ H_{1} ~are ~given~ by\\&
\Pi_{0}=\left(\begin{array}{ccccc}
0 &0&1&0&0\\
0&0&0&1&0\\
-1&0&0&0&0\\
0&-1&0&0&0\\0&0&0&0&2\partial_{t}
 \end{array}\right),\\
&
H_{1}=\frac{1}{2}P^{2}-4p^{2}-\frac{1}{2}\lambda_{1}Q^{2}+\frac{1}{2}qQ^{2}-\frac{1}{8}
q^{3}+\frac{1}{2}cq+\frac{1}{2}\frac{\mu}{Q^{2}}.
 \end{align}
\end{subequations}
The Rosochatius deformation of mKdV equation with self-consistent
source (RD-mKdVESCS) is defined as \cite{y}
\begin{subequations}
\label{eqns:rdmkdv}
 \begin{align}
&v_{t}=\frac{1}{4}(v_{xxx}-6v^{2}v_{x})+\frac{1}{2}(\bar{\varphi_{1}}\bar{\varphi_{2}})_{x},\\
&
\bar{\varphi}_{1x}=v\bar{\varphi_{1}}+\lambda_{1}\bar{\varphi_{2}},~\bar{\varphi_{2x}}=\bar{\varphi_{1}}
-v\bar{\varphi_{2}}+\frac{\mu}{\lambda_{1}\bar{\varphi_{1}}^{3}}.
 \end{align}
\end{subequations}
Let
$$\frac{1}{4}(v_{xx}-2v^{3})+\frac{1}{2}\bar{\varphi_{1}}\bar{\varphi_{2}}=-\bar{c},~v_{t}=-\bar{c}_{x}, $$
\begin{equation} \label{eqns:let}
\bar{q}=v,~\bar{p}=\frac{1}{2}v_{x},~\bar{Q}=\bar{\varphi_{1}},~\bar{P}=\bar{\varphi_{2}},~\bar{R}=
(\bar{Q},\bar{q},\bar{P},\bar{p},\bar{c})^{T},
\end{equation}
then RD-mKdVESCS (\ref{eqns:rdmkdv}) can be written as a $t-$ type
Hamiltonian system
\begin{subequations}
\label{eqns:mkdvh}
 \begin{align}
&\bar{R}_{x}=\left(\begin{array}{c}
\bar{q}\bar{Q}+\lambda_{1}\bar{P}\\
2\bar{p}\\
\bar{Q}-\bar{q}\bar{P}+\frac{\mu}{\lambda_{1}\bar{Q}^{3}}\\
\bar{q}^{3}-\bar{Q}\bar{P}-2\bar{c}\\-\bar{q}_{t}
 \end{array}\right)=\bar{K}_{1}=\bar{\Pi}_{0}\nabla \bar{H}_{1},\\ \nonumber
& where~ \nabla \bar{H}=(\frac{\delta \bar{H}}{\delta
\bar{Q}},\frac{\delta \bar{H}}{\delta \bar{q}},\frac{\delta
\bar{H}}{\delta \bar{P}},\frac{\delta \bar{H}}{\delta
\bar{p}},\frac{\delta \bar{H}}{\delta \bar{c}})^{T}, ~and ~the~ t-type~ Poisson ~operator ~\bar{\Pi}_{0}\\
\nonumber &and ~conserved~ density~ \bar{H}_{1} ~are ~given~ by\\&
\bar{\Pi}_{0}=\left(\begin{array}{ccccc}
0 &0&1&0&0\\
0&0&0&1&0\\
-1&0&0&0&0\\
0&-1&0&0&0\\0&0&0&0&-\frac{1}{2}\partial_{t}
 \end{array}\right),\\
&
\bar{H}_{1}=\bar{q}\bar{P}\bar{Q}+\frac{1}{2}\lambda_{1}\bar{P}^{2}+\bar{p}^{2}
-\frac{1}{2}\bar{Q}^{2}-\frac{1}{4}\bar{q}^{4}+2\bar{c}\bar{q}+\frac{1}{2}\frac{\mu}{\lambda_{1}\bar{Q}^{2}}.
 \end{align}
\end{subequations}
The Miura map relating systems (\ref{eqns:kdvh}) to
(\ref{eqns:mkdvh}), ie $R=M(\bar{R})$, is given by
$$M:~Q=\bar{Q},$$$$q=-\bar{q}^{2}-2\bar{p},$$$$P=\lambda_{1}\bar{P}+\bar{q}\bar{Q},$$
$$p=\frac{1}{4}\bar{q}^{3}-\frac{1}{2}\bar{c}-\frac{1}{4}\bar{Q}\bar{P}+\frac{1}{2}\bar{q}\bar{p},$$
\begin{equation} \label{eqns:map}
c=\bar{H_{1}}-\bar{q_{t}}=-\frac{1}{2}\bar{Q}^{2}-\frac{1}{4}\bar{q}^{4}+\bar{p}^{2}
+2\bar{c}\bar{q}+\frac{1}{2}\lambda_{1}
\bar{P}^{2}+\bar{q}\bar{Q}\bar{P}+\frac{1}{2}\frac{\mu}{\lambda_{1}\bar{Q}^{2}}-\bar{q}_{t}
\end{equation}
which can be proved through direct calculations.\\
Denote $$M'\equiv \frac{DR}{D\bar{R}^{T}}$$ where
$\frac{DR}{D\bar{R}^{T}}$ is the Jacobi matrix consisting of Frechet
derivative of $M$, $M^{'*}$ denotes adjoint of  $M'$. According to
the standard procedure\cite{6}-\cite{9}, applying the map $M$
(\ref{eqns:map}) to the first Hamiltonian structure of RD-mKdVESCS
(\ref{eqns:mkdvh}), we can generate the second Hamiltonian structure
of the RD-KdVESCS (\ref{eqns:kdvh}),
$$\Pi_{1}=M'\bar{\Pi}_{0}M^{'*}$$
\begin{equation} \label{eqns:secondh}
=\left(\begin{array}{ccccc}
0 &0&\lambda_{1}&-\frac{1}{4}Q&P\\
0&0&2Q&-\frac{1}{2}q&-8p+2\partial_{t}\\
-\lambda_{1}&-2Q&0&\frac{1}{4}P&(\lambda_{1}-q)Q+\frac{\mu}{Q^{3}}\\
\frac{1}{4}Q&\frac{1}{2}q&-\frac{1}{4}P&-\frac{1}{8}\partial_{t}&\frac{3}{8}q^{2}-\frac{1}{2}c-\frac{1}{2}Q^{2}
\\-P&8p+2\partial_{t}&(q-\lambda_{1})Q-\frac{\mu}{Q^{3}}&-\frac{3}{8}q^{2}+\frac{1}{2}c+\frac{1}{2}Q^{2}
&q\partial_{t}+\partial_{t}q
 \end{array}\right)
\end{equation}
and the bi-Hamiltonian structure for KdV6 equation or RD-KdVESCS
(\ref{eqns:rdkdv6}) under the transformation (\ref{eqns:wevarphi})
and (\ref{eqns:set}) is given by
\begin{equation}
\label{eqns:bh} R_{x}=\Pi_{0}\frac{\delta H_{1}}{\delta
R}=\Pi_{1}\frac{\delta H_{0}}{\delta R},~H_{0}=c.
\end{equation}
Since the $t-$ type Possion operator $\Pi_{0}$ is invertible, we can
immediately construct a recursion operator
$$\Phi=\Pi_{1}(\Pi_{0})^{-1}$$
\begin{equation} \label{eqns:ro}
=\left(\begin{array}{ccccc}
\lambda_{1} &-\frac{1}{4}Q&0&0&\frac{1}{2}P\partial_{t}^{-1}\\
2Q&-\frac{1}{2}q&0&0&-4p\partial_{t}^{-1}+1\\
0&\frac{1}{4}P&\lambda_{1}&2Q&\frac{1}{2}[(\lambda_{1}-q)Q+\frac{\mu}{Q^{3}}]\partial_{t}^{-1}\\
-\frac{1}{4}P&-\frac{1}{8}\partial_{t}^{-1}&-\frac{1}{4}Q&-\frac{1}{2}q&\frac{1}{2}(\frac{3}{8}q^{2}
-\frac{1}{2}c-\frac{1}{2}Q^{2})\partial_{t}^{-1}
\\(-\lambda_{1}+q)Q-\frac{\mu}{Q^{3}}& -\frac{3}{8}q^{2}
+\frac{1}{2}c+\frac{1}{2}Q^{2}&P&-8p-2\partial_{t}&\frac{1}{2}q+\frac{1}{2}\partial_{t}q\partial_{t}^{-1}
 \end{array}\right)
\end{equation}
which has the hereditary property\cite{6}. Applying $\Phi$ to the
vector field $K_{1}$ (\ref{eqns:kdvh}), we can generate the
hierarchy of Hamiltonian commuting vector fields (symmetries)
\begin{equation} \label{eqns:symm}
K_{n}=\Phi^{n-1}K_{1},
\end{equation}
and obtain a hierarchy of infinite-dimensional integrable
bi-Hamiltonian systems
\begin{equation} \label{eqns:idih}
R_{x}=\Phi^{n-1}K_{1}=K_{n}=\Pi_{0}\nabla H_{n}=\Pi_{1}\nabla
H_{n-1},
\end{equation}
for example
$$K_{2}=\left(\begin{array}{c}
\lambda_{1}P+2pQ+\frac{1}{2}qP \\
2Q\bar{P}+q_{t}\\
-2p\bar{P}+\lambda_{1}^{2}Q-\frac{1}{2}\lambda_{1}Qq+\frac{1}{4}q^{2}Q-Q^{3}-cQ+\frac{\lambda_{1}\mu}{Q^{3}}
+\frac{q\mu}{2Q^{3}}
\\
-\frac{1}{4}P^{2}+p_{t}+\frac{1}{4}qQ^{2}-\frac{1}{4}\lambda_{1}Q^{2}-\frac{1}{4}\frac{\mu}{Q^{2}}\\
2QQ_{t}+c_{t}
 \end{array}\right)$$
with the conserved functional densities $H_{2}$ given by
$$H_{2}=\frac{1}{4}Q^{4}+2pPQ-\frac{1}{8}q^{2}Q^{2}+\frac{1}{4}P^{2}q+\frac{1}{4}\lambda_{1}qQ^{2}
-\frac{1}{2}\lambda_{1}^{2}Q^{2}+\frac{1}{2}\lambda_{1}P^{2}+\frac{\mu
q}{4Q^{2}}+\frac{\mu
\lambda_{1}}{2Q^{2}}-qp_{t}+\frac{1}{2}cQ^{2}+\frac{1}{4}c^{2}.$$

\section{The new solutions of KdV6}
The KdV equation is
\begin{equation} \label{eqns:kdv}
u_{t}=\frac{1}{4}(u_{xxx}+6uu_{x}).
\end{equation}
The lax pair is
\begin{subequations}
\label{eqns:laxpofkdv}
 \begin{align}
& \psi_{xx}+u\psi=\lambda\psi,\\
& \psi_{t}=-\frac{u_{x}}{4}\psi+(\frac{u}{2}+\lambda)\psi_{x}.
 \end{align}
\end{subequations}
Let the functions $\phi_{1},~\phi_{2},\cdots, \phi_{n}$ be n
different solutions of the system (\ref{eqns:laxpofkdv}) with the
corresponding $\lambda=\lambda_{1},~\lambda_{2},\cdots,
\lambda_{n}$. We construct
 two Wronskian determinants from these functions:
\begin{subequations}
\label{eqns:wronsk}
 \begin{align}
&
W_{1}=W(\phi_{1},\cdots,\phi_{1}^{(m_{1})},\phi_{2},\cdots,\phi_{2}^{(m_{2})},\cdots,\phi_{n},
\cdots,\phi_{n}^{(m_{n})}),\\
&
W_{2}=W(\phi_{1},\cdots,\phi_{1}^{(m_{1})},\phi_{2},\cdots,\phi_{2}^{(m_{2})},\cdots,\phi_{n},
\cdots,\phi_{n}^{(m_{n})},\psi),
 \end{align}
\end{subequations}
where $m_{i}\geq 0$ are given numbers and
$\phi_{j}^{(n)}:=\partial_{\lambda}^{n}\phi_{j}(x,\lambda)|_{\lambda=\lambda_{j}}$.
The generalized Darboux transformation of equation (\ref{eqns:kdv})
and  system (\ref{eqns:laxpofkdv}) is given by \cite{vb}
\begin{subequations}
\label{eqns:dt}
 \begin{align}
&
\bar{u}=u+2\partial_{x}^{2}lnW_{1},\\
& \bar{\psi}=\frac{W_{2}}{W_{1}},
 \end{align}
\end{subequations}
 namely, system
(\ref{eqns:laxpofkdv}) is covariant with respect to the action of
(\ref{eqns:dt}). For any initial solution of (\ref{eqns:kdv}),
$\bar{u}$ and  $\bar{\psi}$ are new solution of (\ref{eqns:kdv}) and
(\ref{eqns:laxpofkdv}). Now we will take $u=0$ in what follows.

\subsection{soliton solution}
In (\ref{eqns:wronsk}), let
$n=1,~m_{1}=0,~\lambda=\frac{k^{2}}{4},~\lambda_{1}=\frac{k_{1}^{2}}{4}$
and take
\begin{subequations}
\label{eqns:ss}
 \begin{align}
& \phi_{1}(x,t,k)=cosh\Theta,~\psi_{1}(x,t,k)=sinh\Theta,\\
&\Theta=\frac{k}{2}(x+\frac{1}{4}k^{2}t)+\alpha,~\Theta_{1}=\frac{k_{1}}{2}(x+\frac{1}{4}k_{1}^{2}t)+\alpha
 \end{align}
\end{subequations}
where $\alpha$ is an arbitrary constant. By using
(\ref{eqns:wronsk}) and (\ref{eqns:dt}), we obtain the
single-soliton solution and the corresponding eigenfunction with
$k=k_{1}$ for the KdV equation (\ref{eqns:kdv})
\begin{equation} \label{eqns:os}
\bar{u}=\frac{k_{1}^{2}}{2}sech^{2}\Theta_{1},
\end{equation}
\begin{equation} \label{eqns:eigen}
\bar{\psi}_{1}(x,t,k_{1})=\frac{\beta k_{1}}{2}sech\Theta_{1},
\end{equation}
where $\beta$ is an arbitrary constant as well.

Since  KdV6 equation (\ref{eqns:nkdv6}) can be considered to be
 KdV equation (\ref{eqns:kdv}) with non-homogeneous terms and $w$ is related to the
 square of eigenfunction by (\ref{eqns:wevarphi}), we may apply the
method of variation of constant to find the solution  of
Eq.(\ref{eqns:nkdv6}) by using the solution $\bar{u}$ of
Eq.(\ref{eqns:kdv}) and corresponding eigenfunction
$\bar{\psi}_{1}$. Taking $\alpha$ and $\beta$ in (28b) and
(\ref{eqns:eigen}) to be time-dependent functions $\alpha(t)$ and
$\beta(t)$ and using (\ref{eqns:wevarphi}),  and requiring that
\begin{subequations}
\label{eqns:uw}
 \begin{align}
&
u=\frac{k_{1}^{2}}{2}sech^{2}\bar{\Theta}_{1},\\
& w=\bar{\psi}_{1}^{2}(x,t,k_{1})=\frac{\beta(t)^{2}k_{1}^{2}}{4}sech^{2}\bar{\Theta}_{1},\\
&\bar{\Theta}_{1}=\frac{k_{1}}{2}(x+\frac{1}{4}k_{1}^{2}t)+\alpha(t),
 \end{align}
\end{subequations}
satisfy the Eq.(\ref{eqns:nkdv6}). We find that $\alpha(t)$ can be
an arbitrary function of $t$ and
\begin{equation} \label{eqns:beta}
\beta(t)^{2}=-\frac{4\alpha'(t)}{k_{1}}.
\end{equation}
So the single-soliton solution of KdV6 equation (\ref{eqns:nkdv6})
is given by
\begin{subequations}
\label{eqns:uw}
 \begin{align}
&
u=\frac{k_{1}^{2}}{2}sech^{2}\bar{\Theta}_{1},\\
& w=-\alpha'(t)k_{1}sech^{2}\bar{\Theta}_{1}.
 \end{align}
\end{subequations}
Its shape is shown in figure 1. Notice that $\bar{\Theta}_{1}$
contains an arbitrary t-function $\alpha(t)$. This implies that the
insertion of sources into KdV equation may cause the variation of
the speed of the soliton solution. So the dynamics of solution of
KdV6 equation turns out to be much richer than that of solution of
KdV equation.

\vskip 80pt
\begin{center}
\begin{picture}(35,25)
\put(-150,0){\resizebox{!}{3.3cm}{\includegraphics{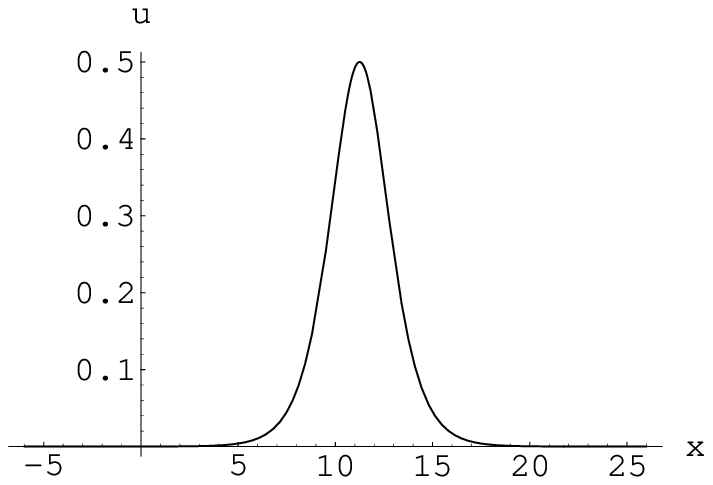}}}
\put(60,0){\resizebox{!}{3.3cm}{\includegraphics{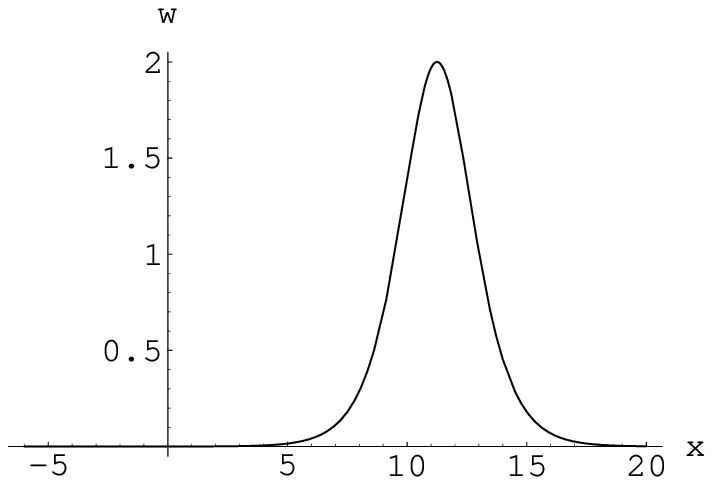}}}
\end{picture}
\end{center}
\begin{center}
\begin{minipage}{ 12cm}{\footnotesize
~~~~~~~~~~~~~~~(a)~~~~~~~~~~~~~~~~~~~~~~~~~~~~~~~~~~~~~~~~~~~~~~
~~~~~~~~~~~~~~(b)\\
{\bf Figure 1.} The shape of single soliton solution  for $u$ and
$w$ when $\alpha(t)=-2t,~k_{1}=1,~t=3.$ }
\end{minipage}
\end{center}

\subsection{The first and second order of positon solution}
In (\ref{eqns:wronsk}), set
$n=1,~m_{1}=1,~\lambda=-\frac{k^{2}}{4},~\lambda_{1}=-\frac{k_{1}^{2}}{4}$
and take
\begin{subequations}
\label{eqns:p1}
 \begin{align}
& \phi_{1}(x,t,k)=sin\Theta,~\psi_{1}(x,t,k)=cos\Theta,\\
&\Theta=\frac{k}{2}(x+x_{1}{(k)}-\frac{1}{4}k^{2}t)-\frac{1}{8}(k-k_{1})\alpha,
 \end{align}
\end{subequations}
where $x_{1}(k)$ is a function that is analytic in the vicinity of
the point $k$ and has real Taylor expansion coefficients.  By using
(\ref{eqns:wronsk}) , (\ref{eqns:dt}) and (\ref{eqns:p1}), we obtain
first order of  the one-positon solution and the corresponding
eigenfunction with $k=k_{1}$ for the KdV equation
(\ref{eqns:kdv})\cite{vb}
\begin{subequations}
\label{eqns:p2}
 \begin{align}
& \bar{u}=\frac{-16k_{1}^{2}sin\Theta_{1}(8sin\Theta_{1}+k_{1}\gamma
cos\Theta_{1})}
{(4sin2\Theta_{1}+k_{1}\gamma)^{2}},\\
& \bar{\psi}_{1}(x,t,k_{1})=-\frac{4\beta k_{1}^{2}sin\Theta_{1}}{4sin2\Theta_{1}+k_{1}\gamma},\\
&\Theta_{1}=\frac{k_{1}}{2}(x+x_{1}{(k_{1})}-\frac{1}{4}k_{1}^{2}t),\\
&\gamma=-8\partial_{k}\Theta|_{k=k_{1}}=3k_{1}^{2}t-4(x+x_{2}(k_{1}))+\alpha,~x_{2}(k_{1})=[x_{1}
+4k\partial_{k}x_{1}(k)]_{k=k_{1}}
 \end{align}
\end{subequations}
where $\alpha,~\beta$ are arbitrary constants.
Similarly, by using (\ref{eqns:wevarphi})  and the method of
variation of constant we present first order of the one-positon
solution for the KdV6 equation (\ref{eqns:nkdv6})
\begin{subequations}
\label{eqns:p4}
 \begin{align}
& u=\frac{-16k_{1}^{2}sin\Theta_{1}(8sin\Theta_{1}+k_{1}\bar{\gamma}
cos\Theta_{1})}
{(4sin2\Theta_{1}+k_{1}\bar{\gamma})^{2}},\\
& w=-\frac{16k_{1}^{2}\alpha'(t)sin^{2}\Theta_{1}}{(4sin2\Theta_{1}+k_{1}\bar{\gamma})^{2}},\\
&\bar{\gamma}=3k_{1}^{2}t-4(x+x_{2}(k_{1}))+\alpha(t).
 \end{align}
\end{subequations}
(\ref{eqns:p4}) implies that for fixed $t$ and
$x\rightarrow\pm\infty$, we have the asymptotic estimate
\begin{subequations}
\label{eqns:p5}
 \begin{align}
& u=\frac{2k_{1}}{x}sin2\Theta_{1}[1+O(x^{-1})],\\
& w=-\frac{k_{1}\alpha'(t)}{x^{2}}sin^{2}\Theta_{1}[1+O(x^{-1})].
 \end{align}
\end{subequations}
If $x$ is fixed and $t\rightarrow\pm\infty$, the solution has the
asymptotic behavior
\begin{subequations}
\label{eqns:p6}
 \begin{align}
& u=-\frac{8sin2\Theta_{1}}{3k_{1}t}[1+O(t^{-1})],\\
&
w=-\frac{16\alpha'(t)}{9k_{1}^{4}t^{2}}sin^{2}\Theta_{1}[1+O(t^{-1})].
 \end{align}
\end{subequations}
A positon solution as a function of $x,~u,$ and $w$ have a
second-order pole. This pole is situated at the point $x=x_{0}(t)$
which oscillates around the point $x_{as}(t)$ with the amplitude
$\frac{4}{k_{1}}$, where
$x_{as}(t)=-\frac{3}{4}k_{1}^{2}t-x_{2}(k_{1})-\frac{\alpha(t)}{4}$.
The exact position of this pole can be determined by solving the
following equation with
$\delta=k_{1}\bar{\gamma},$$$\delta=-4sin\frac{1}{8}[\delta-4k_{1}^{3}t+4k_{1}(x_{1}-x_{2})-k_{1}\alpha(t)].$$
So the positon solution of KdV6 equation (\ref{eqns:nkdv6}) is
long-range analogue of soliton and is slowly decreasing, oscillating
solution. The shape and motion of the single positon is shown in
figure 2.

\vskip 80pt
\begin{center}
\begin{picture}(25,25)
\put(-220,0){\resizebox{!}{3.0cm}{\includegraphics{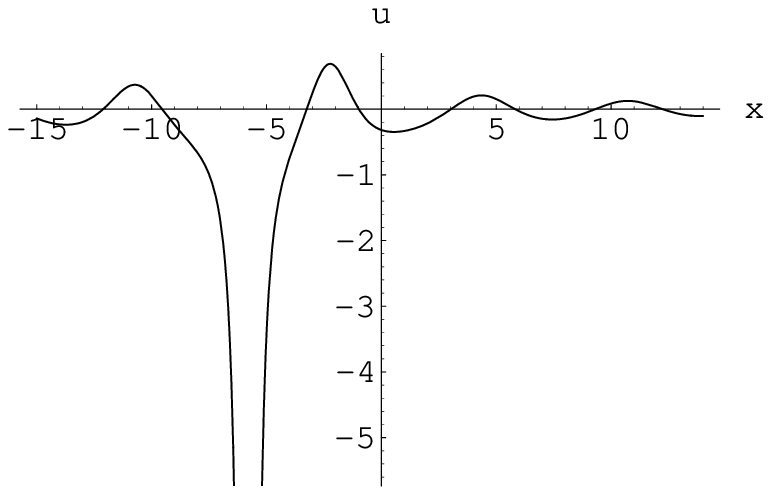}}}
\put(-70,0){\resizebox{!}{3.0cm}{\includegraphics{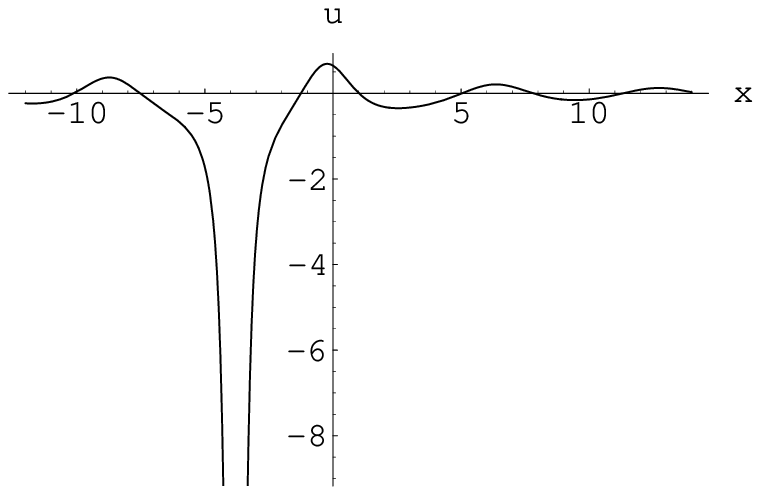}}}
\put(90,0){\resizebox{!}{3.0cm}{\includegraphics{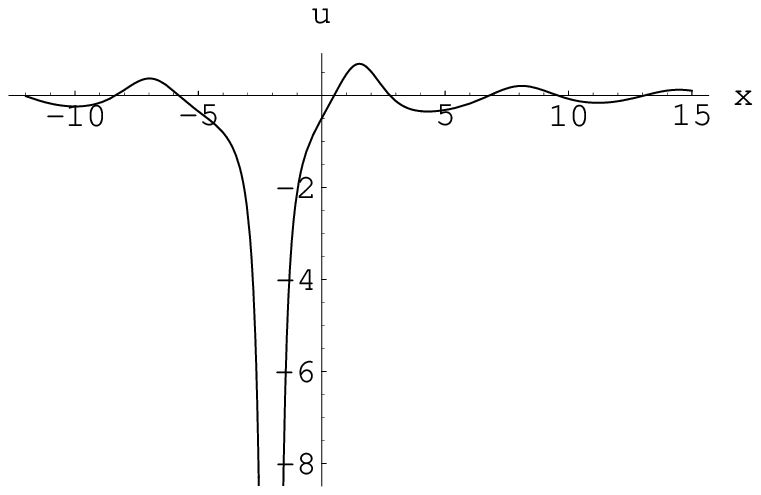}}}
\end{picture}
\end{center}
\begin{center}
\begin{minipage}{ 14cm}{\footnotesize
~~~~t=-5~~~~~~~~~~~~~~~~~~~~~~~~~~~~~~~~~~~~~t=3~~~~~~~~~~~~~~~~
~~~~~~~~~~~~~~~~~~~~~~~~~~~~t=10}
\end{minipage}
\end{center}
\vskip 80pt
\begin{center}
\begin{picture}(25,25)
\put(-220,0){\resizebox{!}{3.0cm}{\includegraphics{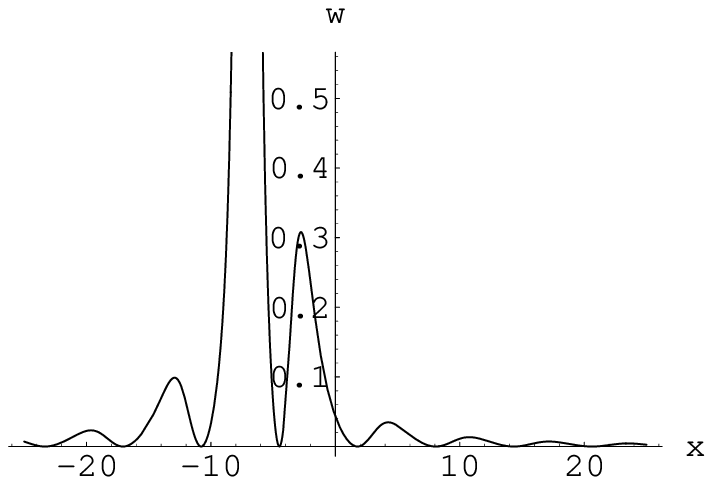}}}
\put(-70,0){\resizebox{!}{3.0cm}{\includegraphics{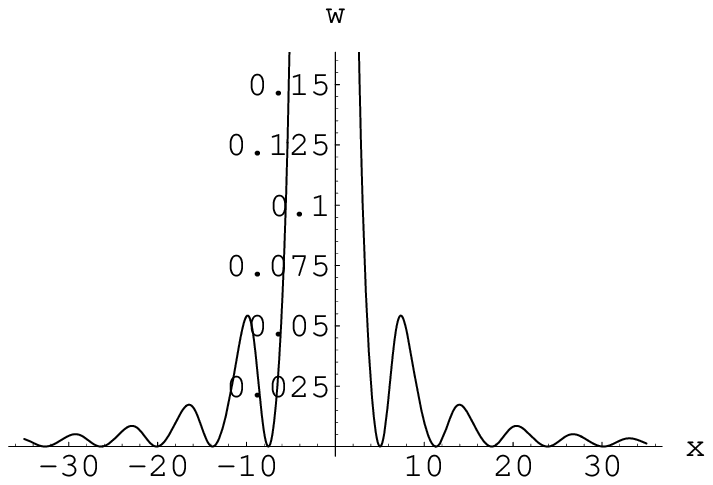}}}
\put(90,0){\resizebox{!}{3.0cm}{\includegraphics{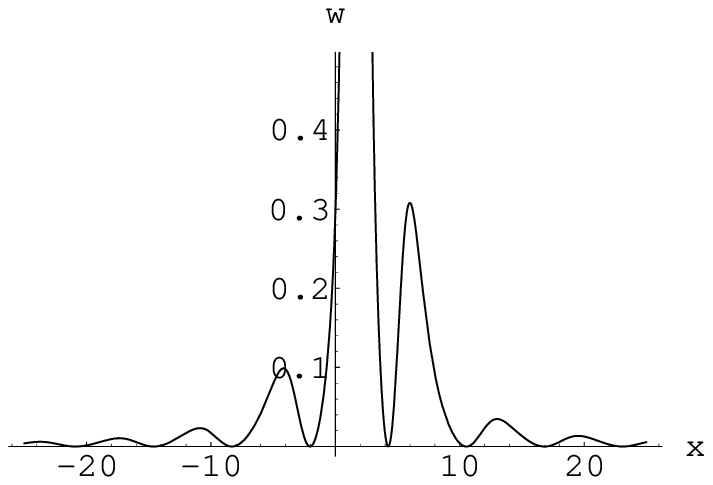}}}
\end{picture}
\end{center}
\begin{center}
\begin{minipage}{ 14cm}{\footnotesize
~~~~t=-10~~~~~~~~~~~~~~~~~~~~~~~~~~~~~~~~~~~~~t=3~~~~~~~~~~~~~~~~
~~~~~~~~~~~~~~~~~~~~~~~~~~~~t=25\\
{\bf Figure 2.} The shape and motion of one-positon solution  for
$u$ and $w$ when $\alpha(t)=-2t,~x_{1}(k_{1})=2k_{1},~k_{1}=1.$ }
\end{minipage}
\end{center}
In order to find the second order of one-positon solution, we take
$\Theta$ in (34a) to be
\begin{equation}\Theta=\frac{k}{2}(x+x_{1}{(k)}-\frac{1}{4}k^{2}t)-\frac{1}{8}(k-k_{1})^{2}\alpha,
\end{equation}
we have
\begin{subequations}
\label{eqns:p7}
 \begin{align}
&\nonumber
W_{1}=W(\phi_{1},\partial_{k}\phi_{1},\partial_{k}^{2}\phi_{1})|_{k=k_{1}}=\frac{1}{128}\{-32sin^{2}\Theta_{1}cos\Theta_{1}+k_{1}^{2}
\gamma^{2}cos\Theta_{1}+[12k_{1}^{2}\nu\\
&~~~~~~~~~~~~~~~~~~~~~~~~~~~~~~~~~~~~~~-4k_{1}
(4x+4x_{1}(k_{1})+k_{1}\alpha-2k_{1}^{2}x_{1}''(k_{1}))]sin\Theta_{1}\},\\
&W_{2}=W(\phi_{1},\partial_{k}\phi_{1},\partial_{k}^{2}\phi_{1},\psi_{1})|_{k=k_{1}}=-\frac{1}{64}
k_{1}^{3}(4sin2\Theta_{1}+k_{1}\gamma),\\
&
\Theta_{1}=\frac{k_{1}}{2}(x+x_{1}{(k_{1})}-\frac{1}{4}k_{1}^{2}t),~\gamma=
-8\partial_{k}\Theta|_{k=k_{1}}=3k_{1}^{2}t-4(x+x_{2}(k_{1})),\\
&
\nu=-4\partial_{k}^{2}\Theta|_{k=k_{1}}=3k_{1}t-4\partial_{k}x_{2}{(k)}|_{k=k_{1}}+\alpha.
 \end{align}
\end{subequations}
By the similar process , by taking $\alpha=\alpha(t)$ in (40d) we
obtain the second-order of the one-positon solution from (7) and
(27)
\begin{subequations}
\label{eqns:p8}
 \begin{align}
& u=2\partial_{x}^{2}lnW_{1},\\
& w=-\frac{2\alpha'(t)}{k_{1}^{3}}(\frac{W_{2}}{W_{1}})^{2}.
 \end{align}
\end{subequations}
Its shape  is shown in figure 3.

\vskip 80pt
\begin{center}
\begin{picture}(35,25)
\put(-150,0){\resizebox{!}{3.3cm}{\includegraphics{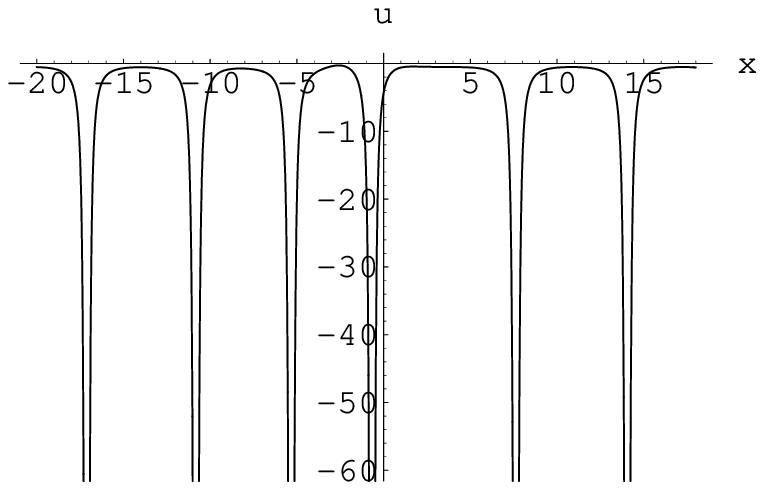}}}
\put(60,0){\resizebox{!}{3.3cm}{\includegraphics{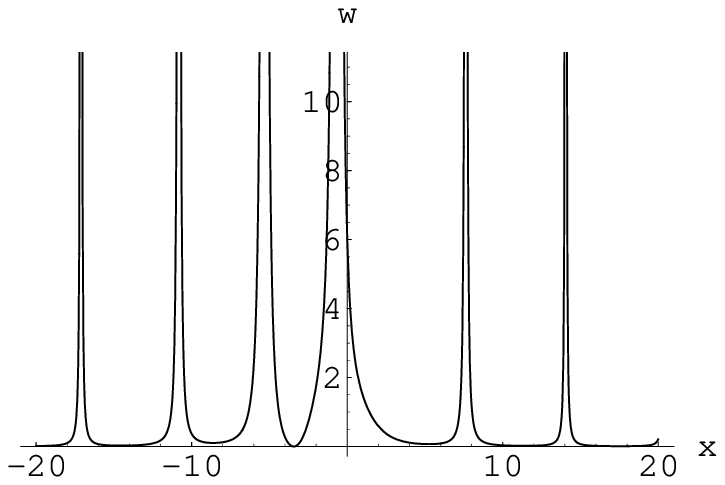}}}
\end{picture}
\end{center}
\begin{center}
\begin{minipage}{ 12cm}{\footnotesize
~~~~~~~~~~~~~~~(a)~~~~~~~~~~~~~~~~~~~~~~~~~~~~~~~~~~~~~~~~~~~~~~
~~~~~~~~~~~~~~(b)\\
{\bf Figure 3.} The shape of second-order positon solution  for $u$
and $w$ when $x_{1}(k_{1})=2k_{1},~\alpha(t)=-2t^{2},~k_{1}=1,~t=2.$
}
\end{minipage}
\end{center}

\subsection{The first and second order of negaton solution}
In (\ref{eqns:wronsk}), set
$n=1,~m_{1}=1,~\lambda=\frac{k^{2}}{4},~\lambda_{1}=\frac{k_{1}^{2}}{4}$
and take
\begin{subequations}
\label{eqns:n1}
 \begin{align}
& \phi_{1}(x,t,k)=sinh\Theta,~\psi_{1}(x,t,k)=cosh\Theta,\\
&\Theta=\frac{k}{2}(x+x_{1}{(k)}+\frac{1}{4}k^{2}t)+\frac{1}{8}(k-k_{1})\alpha.
 \end{align}
\end{subequations}
We obtain the first order of  one-negaton solution and the
corresponding eigenfunction with $k=k_{1}$ for the KdV equation
(\ref{eqns:kdv})
\begin{subequations}
\label{eqns:n2}
 \begin{align}
&
\bar{u}=\frac{-16k_{1}^{2}sinh\Theta_{1}(8sinh\Theta_{1}-k_{1}\gamma
cosh\Theta_{1})}
{(4sinh2\Theta_{1}-k_{1}\gamma)^{2}},\\
& \bar{\psi}_{1}(x,t,k_{1})=-\frac{4\beta k_{1}^{2}sinh\Theta_{1}}{4sinh2\Theta_{1}-k_{1}\gamma},\\
&\Theta_{1}=\frac{k_{1}}{2}(x+x_{1}{(k_{1})}+\frac{1}{4}k_{1}^{2}t),\\
&\gamma=8\partial_{k}\Theta|_{k=k_{1}}=3k_{1}^{2}t+4(x+x_{2}(k_{1}))+\alpha,
 \end{align}
\end{subequations}
where $\alpha,~\beta$ are arbitrary constants.

\vskip 80pt
\begin{center}
\begin{picture}(25,25)
\put(-220,0){\resizebox{!}{3.0cm}{\includegraphics{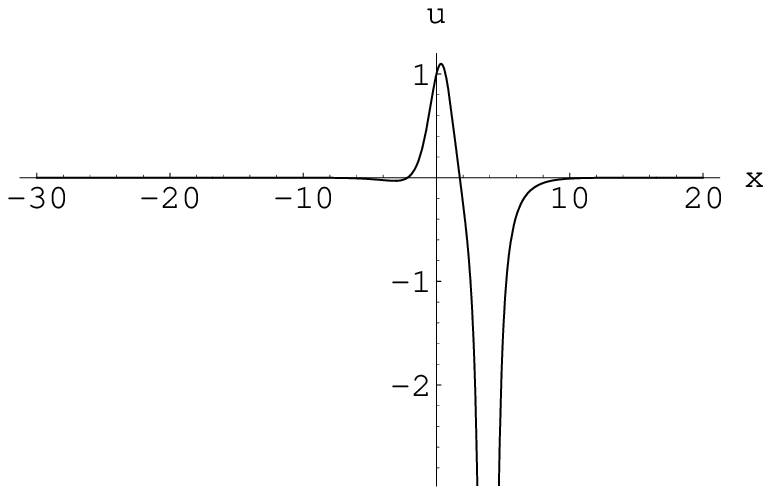}}}
\put(-70,0){\resizebox{!}{3.0cm}{\includegraphics{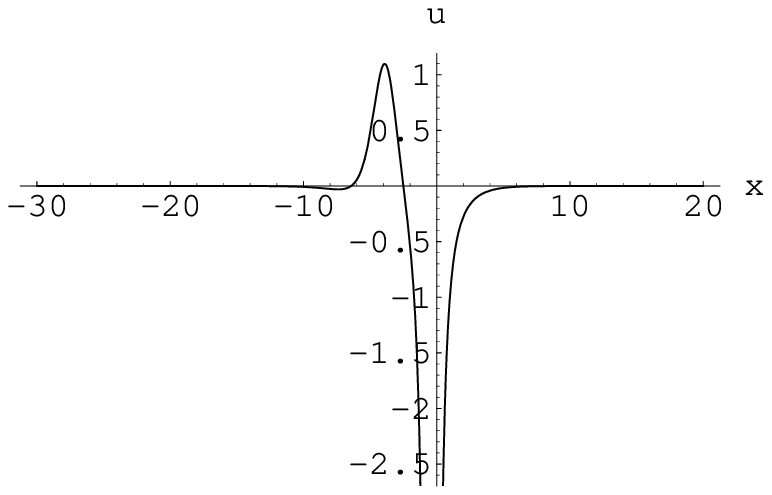}}}
\put(90,0){\resizebox{!}{3.0cm}{\includegraphics{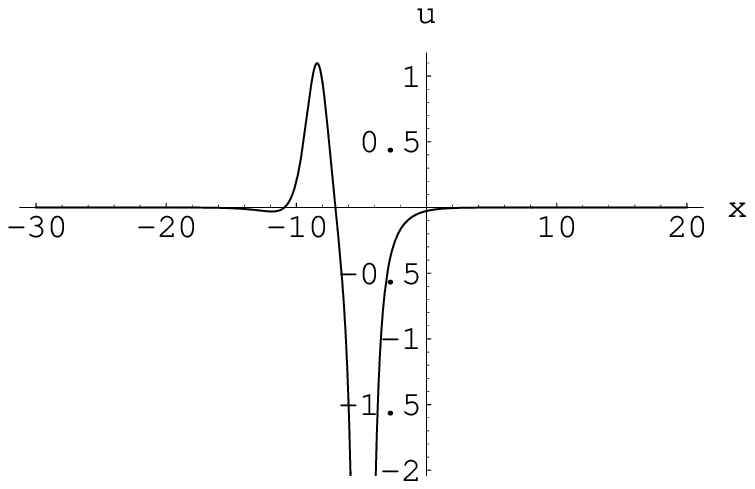}}}
\end{picture}
\end{center}
\begin{center}
\begin{minipage}{ 14cm}{\footnotesize
~~~~~~t=-15~~~~~~~~~~~~~~~~~~~~~~~~~~~~~~~~~~~~~t=2~~~~~~~~~~~~~~~~
~~~~~~~~~~~~~~~~~~~~~~~~~~~~t=20}
\end{minipage}
\end{center}
\vskip 80pt
\begin{center}
\begin{picture}(25,25)
\put(-220,0){\resizebox{!}{3.0cm}{\includegraphics{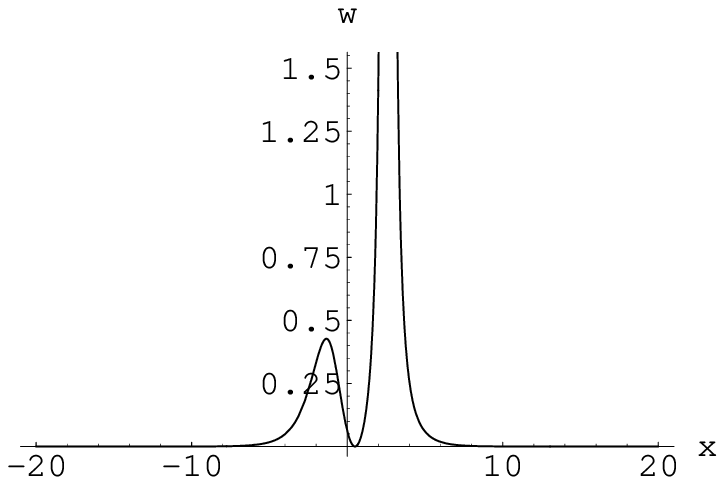}}}
\put(-70,0){\resizebox{!}{3.0cm}{\includegraphics{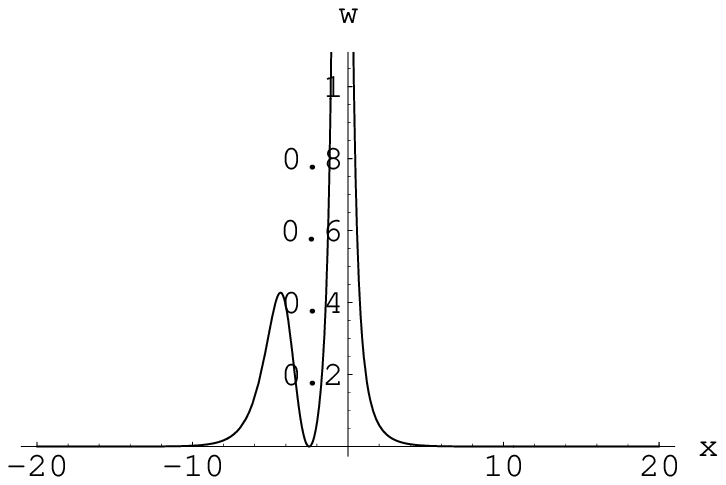}}}
\put(90,0){\resizebox{!}{3.0cm}{\includegraphics{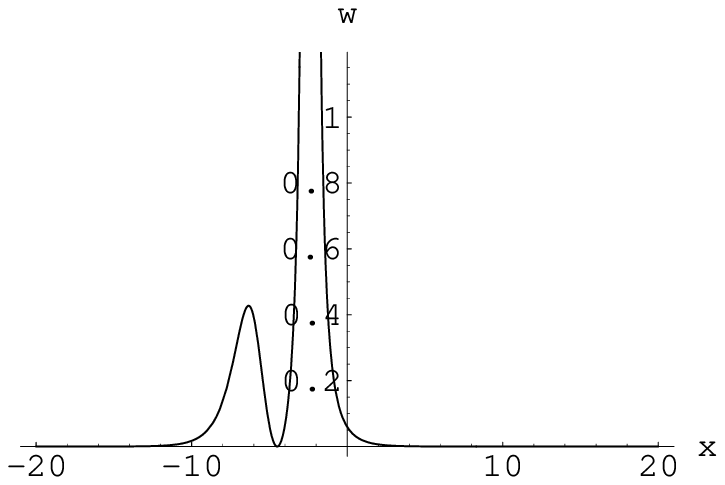}}}
\end{picture}
\end{center}
\begin{center}
\begin{minipage}{ 14cm}{\footnotesize
~~~~t=-10~~~~~~~~~~~~~~~~~~~~~~~~~~~~~~~~~~~~~t=2~~~~~~~~~~~~~~~~
~~~~~~~~~~~~~~~~~~~~~~~~~~~~t=10\\
{\bf Figure 4.} The shape and motion of one-negaton solution  for
$u$ and $w$ when $\alpha(t)=-2t,~x_{1}(k_{1})=2k_{1},~k_{1}=1.$ }

Similarly, by using (\ref{eqns:wevarphi}) and the method of
variation of constant we present the first order of  one-negaton
solution for the KdV6 equation (\ref{eqns:nkdv6})
\end{minipage}
\end{center}
\begin{subequations}
\label{eqns:n4}
 \begin{align}
&
u=\frac{-16k_{1}^{2}sinh\Theta_{1}(8sinh\Theta_{1}-k_{1}\bar{\gamma}
cosh\Theta_{1})}
{(4sinh2\Theta_{1}-k_{1}\bar{\gamma})^{2}}\\
& w=-\frac{16k_{1}^{2}\alpha'(t)sinh^{2}\Theta_{1}}{(4sinh2\Theta_{1}-k_{1}\bar{\gamma})^{2}},\\
&\bar{\gamma}=3k_{1}^{2}t+4(x+x_{2}(k_{1}))+\alpha(t).
 \end{align}
\end{subequations}
Similarly, negaton solution of (\ref{eqns:nkdv6}) have second-order
pole. The shape and motion of the negaton is shown in figure 4.

Now we take $\Theta$ in (42a) to be
\begin{equation}\Theta=\frac{k}{2}(x+x_{1}{(k)}+\frac{1}{4}k^{2}t)+\frac{1}{8}(k-k_{1})^{2}\alpha,
\end{equation}
we have
\begin{subequations}
\label{eqns:n5}
 \begin{align}
&\nonumber
W_{1}=W(\phi_{1},\partial_{k}\phi_{1},\partial_{k}^{2}\phi_{1})|_{k=k_{1}}=\frac{1}{128}\{32sinh^{2}\Theta_{1}cosh\Theta_{1}-k_{1}^{2}
\gamma^{2}cosh\Theta_{1}+[12k_{1}^{2}\nu
\\
&~~~~~~~~~~~~~~~~~~~~~~~~~~~~~~~~~~~~~-4k_{1}(-4x-4x_{1}(k_{1})+k_{1}\alpha+2k_{1}^{2}x_{1}''(k_{1}))]sinh\Theta_{1}\},\\
&W_{2}=W(\phi_{1},\partial_{k}\phi_{1},\partial_{k}^{2}\phi_{1},\psi_{1})|_{k=k_{1}}=\frac{1}{64}k_{1}^{3}(-4sin2\Theta_{1}+k_{1}\gamma),\\
&
\Theta_{1}=\frac{k_{1}}{2}(x+x_{1}{(k_{1})}+\frac{1}{4}k_{1}^{2}t),~\gamma=
8\partial_{k}\Theta|_{k=k_{1}}=3k_{1}^{2}t+4(x+x_{2}(k_{1})),\\
&
\nu=4\partial_{k}^{2}\Theta|_{k=k_{1}}=3k_{1}t+4\partial_{k}x_{2}{(k)}|_{k=k_{1}}+\alpha.
 \end{align}
\end{subequations}
Similarly, we obtain the second-order of negaton solution from (7)
and (27)
\begin{subequations}
\label{eqns:n6}
 \begin{align}
& u=2\partial_{x}^{2}lnW_{1},\\
& w=\frac{2\alpha'(t)}{k_{1}^{3}}(\frac{W_{2}}{W_{1}})^{2}.
 \end{align}
\end{subequations}
 Its shape  is shown in figure 5.

\vskip 80pt
\begin{center}
\begin{picture}(35,25)
\put(-150,0){\resizebox{!}{3.3cm}{\includegraphics{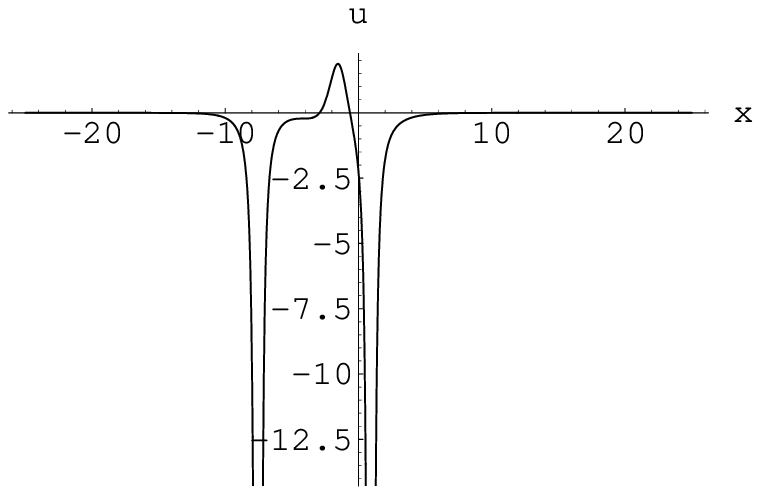}}}
\put(60,0){\resizebox{!}{3.3cm}{\includegraphics{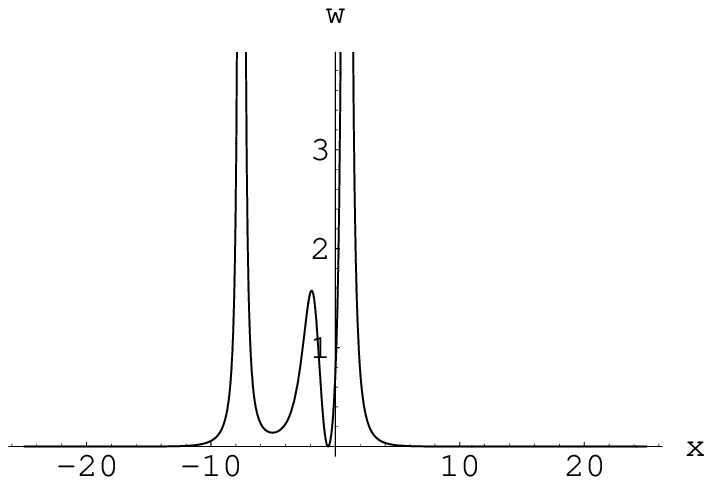}}}
\end{picture}
\end{center}
\begin{center}
\begin{minipage}{ 12cm}{\footnotesize
~~~~~~~~~~~~~~~(a)~~~~~~~~~~~~~~~~~~~~~~~~~~~~~~~~~~~~~~~~~~~~~~
~~~~~~~~~~~~~~(b)\\
{\bf Figure 5.} The shape of second-order negaton solution  for $u$
and $w$ when $x_{1}(k_{1})=2k_{1},~\alpha(t)=2t^{2},~k_{1}=1,~t=5.$
}
\end{minipage}
\end{center}

\section*{Acknowledgments}

This work is supported by National Basic Research Program of China
(973 Program) (2007CB814800) and China Postdoctoral Science
Foundation funded project(20080430420).


\def\cprime{$'$} \def\cprime{$'$} \def\cprime{$'$} \def\cprime{$'$}
  \def\cprime{$'$}

\end{document}